\documentclass[letterpaper, 10 pt, conference]{ieeeconf}  

\IEEEoverridecommandlockouts                              

\usepackage{amsmath} 
\usepackage{amssymb}  
\usepackage{caption, subcaption, subfloat}
\usepackage{relsize}  

\usepackage{xcolor}

\usepackage{circuitikz}
\usepackage{pgfplotstable} 
\usepackage{pgfplots}  
\pgfplotsset{compat=1.16}
\usepgfplotslibrary{fillbetween}
\usepgfplotslibrary{external}
\usetikzlibrary{backgrounds}  

\title{\LARGE \bf
Quantized Deep Path-following Control on a Microcontroller
}

\author{Pablo Zometa$^{1}$ 
 and Timm Faulwasser$^{2}$ 
\thanks{$^{1}$PZ
is with the Faculty of Engineering,
        German International University in Berlin, Germany. 
        {\tt\small pablo.zometa@giu-berlin.de}}
\thanks{$^{2}$TF is with the
Institute of Energy Systems, Energy Efficiency and Energy Economics,
TU Dortmund, Germany. 
        {\tt\small timm.faulwasser@ieee.org}}
}

\begin{document}

\maketitle
\thispagestyle{empty}
\pagestyle{empty}

\newcommand{\ub}[1]{\ensuremath{\underline{\textbf{#1}}}}
\newcommand{\vb}[1]{\ensuremath{\textbf{#1}}}
\newcommand{\inset}[1]{\in \mathcal{#1}}
\newcommand{\set}[1]{\mathcal{#1}}
\newcommand{\inRpow}[1]{\in \mathbb{R}^{#1}}
\newcommand{\inRpownm}[2]{\in \mathbb{R}^{#1 \times #2}}
\newcommand{\Rpown}[1]{\mathbb{R}^{#1}}
\newcommand{\Rpownm}[2]{\mathbb{R}^{#1 \times #2}}
\newcommand{\totalderiv}[2]{\frac{\textrm{d} #1}{\textrm{d} #2}}
\newcommand{\partderiv}[2]{\frac{\partial #1}{\partial #2}}
\newcommand{\Tr}{\ensuremath{\mathsf{\mathsmaller T}}} 
\newcommand{\TF}[1]{\textcolor{red}{#1}}

\begin{abstract}
    Model predictive Path-Following Control (MPFC)  is a viable option for
    motion systems in many application domains. However, despite
    considerable progress on tailored numerical methods for predictive
    control, the real-time implementation of predictive control and MPFC on
    small-scale autonomous platforms with low-cost embedded hardware
    remains challenging.  While usual stabilizing MPC formulations lead to
    static feedback laws, the MPFC feedback turns out
    to be dynamic as the path parameter acts as an internal controller
    variable. In this paper, we leverage deep learning to implement
    predictive path-following control on microcontrollers. We show that
    deep neural networks can approximate the dynamic MPFC feedback law
    accurately. Moreover, we illustrate and tackle the challenges that
    arise if the target platform employs limited precision arithmetic.
    Specifically, we draw upon a post-stabilization with an additional
    feedback law to attenuate undesired quantization effects. Simulation
    examples underpin the efficacy of the proposed approach.
\end{abstract}
\section{Introduction}

Nonlinear Model Predictive Control (NMPC) is a 
control method that can handle nonlinear system dynamics as well as input and state constraints. In its base variant NMPC for setpoint stabilization yields a static feedback law. 
Another variant is
Model predictive Path-Following Control (MPFC), 
which has been successfully applied to motion control of robots
to precisely follow a geometric reference path
\cite{fukushima2020model, faulwasser2016implementation, mehrez2017predictive}. In MPFC the considered reference is a geometric path and timing along the path is computed at the run-time of the controller. Hence and in contrast to NMPC for setpoint stabilization, the MPFC is a dynamic feedback strategy as the reference position is an internal controller memory~\cite{faulwasser2015nonlinear}.

An often cited disadvantage of NMPC is its high computational cost,
which significantly limits its use in low-cost computing hardware 
like MicroController Units (MCU).
The \textrm{Optimization Engine} (OpEn) \cite{open2020} 
and \textrm{acados} \cite{Verschueren2021}, two popular state-of-the-art NMPC solvers,
can efficiently run on embedded
hardware like a Raspberry Pi (a single-board computer).
However, at the time of this writing,
none of them can run out of the box on 32-bit MCUs.

To overcome the high computational demands of NMPC, 
the use of deep neural networks as a way to 
quickly find an approximate solution to the NMPC problem 
has been proposed \cite{parisini1995receding}, \cite{lucia2018}, \cite{kumar2018deep}.
In particular, \cite{lucia2018} explores a 
robust multi-stage NMPC on an MCU
using a Deep Neural Network (DNN) using single-precision floating-point arithmetic during network inference.

Moreover, to further increase the efficiency of DNNs, 
the use of quantization---i.e., storing the network parameters using fixed-point 
representation instead of floating point---has been explored
\cite{gholami2021survey}.
Compared to a regular DNN, a quantized DNN executes much faster, 
requires less memory, and is more energy efficient---there is the downside of some loss of numerical accuracy
\cite{gholami2021survey}.

The  present paper investigates the use of quantized deep neural networks
for model predictive path-following control of mobile robots. 
Our main contribution is two-fold: first, we propose a way to generate 
the training set  that takes into account the path to be followed, and second, we extend the DNN with a simple controller to make up for errors introduced by the quantized DNN approximation.

Using the proposed approach with hardware-in-the-loop simulations 
running on an MCU, we show that
a quantized deep neural network requiring less than $5$ kB of storage memory
achieves a good path following performance while being
several orders of magnitude faster than OpEn.

The remainder of the paper is organized as follows:
Section \ref{sec:mpfc_robot} recalls MPFC 
applied to a mobile robot. Section \ref{sec:qdnn} discusses 
quantized DNNs. Section \ref{sec:qdnn-mpfc}
introduces an approach to efficiently approximate the MPFC problem 
using quantized DNN, followed by the results (Section \ref{sec:results}) and conclusions
(Section \ref{sec:concl}).

\section{Path following control of a mobile robot}
\label{sec:mpfc_robot}

This section summarizes 
the main idea of MPFC
according to \cite{faulwasser2009nonlinear}, and its application 
to differential drive robots \cite{mehrez2017predictive}.

\subsection{System Description}
\begin{figure}[t]
\centering
\begin{tikzpicture}
    \begin{scope}
            \draw[thick, ->] (0,0) -- (3.3,0)  node[right, text width=1em] {$X$};
            \draw[thick, ->] (0,0) -- (0,3.5)  node[above, text width=1em] {$Y$};
        \end{scope}

    \begin{scope}[rotate around={-45:(1.5,1.5)},draw=black]
    \draw[fill=gray!40, rounded corners] (1, 1) rectangle (1.2, 2) node[above left] {$L$};
    \draw[fill=gray!40, rounded corners] (2.4, 1) rectangle (2.6, 2) node[below] {$R$};
    \draw[fill=white, rounded corners] (1.2, 1) rectangle (2.4, 2.6);
    \draw[|latex-{latex}{|}] (1.1, 0.8) -- (1.8, 0.8) node[xshift=-0.4cm] {$\ell$};
    \draw[|latex - {latex}{|}] (1.8, 0.8) -- (2.5, 0.8) node[xshift=-0.4cm] {$\ell$};

    \node at (1.8, 2.4) (piv) {};
    \node at (1.8, 1.5) (cm) {};
    \draw[fill=black] (cm) circle (0.05);
    \draw[very thick, -latex] (cm.center) -- ++(0, 0.7) node[xshift=-0.3cm] {$s$};
    \draw[thick, ->] (cm.center) -- ++(0, 1.8) node[right] {$\hat X$};
    \draw[thick, ->] (cm.center) -- ++(-1.5, 0) node[above] {$\hat Y$};
    \draw[thick, dashed] (cm.center) -- ++(1,1) node (theta) {};
    \node at (theta.north) [above right] (tt) {$\varphi$};
    \draw[thick,-latex] (theta) arc [radius=1.41, start angle=45, end angle=90];
    \end{scope}

    \draw[dashed] (cm) -- ++(-1.8, 0)  node[left] {$q_y$};
    \draw[dashed] (cm) -- ++(0, -1.4)  node[below] {$q_x$};
    \node at (cm) [xshift=+0.15cm, yshift=-0.25cm] {$q$};
    
\end{tikzpicture}
\begin{tikzpicture}
    \begin{axis}[
        xticklabel style={
              /pgf/number format/precision=3,
              /pgf/number format/fixed}, 
        xlabel={$X$},
        ylabel={$Y$},
        ylabel style={rotate=-90, at={(0.1,1.1)}},
        xlabel style={at={(1.15, 0.1)}},
        xmin=-2, xmax=2,
        height=4cm, width=4.0cm]
        \addplot [no marks,
        ] table [x=pos-x, y=pos-y, col sep=comma] {figs/pathparam.csv};
    \end{axis}
\end{tikzpicture}
\caption{Left: differential drive robot and its coordinate systems. 
    Right: the path at scale, an ellipse. The robot's left and right
wheels are marked $L$ and $R$, respectively.}
\label{fig:diffdrvrbt}
\end{figure}
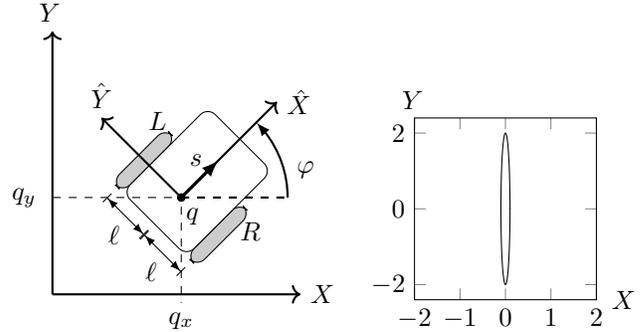

Fig. \ref{fig:diffdrvrbt} shows
a schematic of a differential drive robot. 
The global (inertial) frame is defined by the axes $XY$,
whereas the local frame attached to the robot is defined by the axes $\hat X \hat Y$.
The position of the robot in the global frame is represented by the 
Cartesian coordinates of point $q$ (the origin of the local frame). The robot's pose $\xi$
in the inertial frame 
is represented by its Cartesian position $q=[q_x \ q_y]^\Tr$ and 
orientation $\varphi$,
that is $\xi = [q_x \ q_y \ \varphi]^\Tr$.
We represent the robot dynamics as
the rate of change of the pose in terms of the
robot's forward speed $s$, and its angular velocity $\omega$:
\begin{equation}
\dot \xi = f(\xi,u) = \begin{bmatrix}
    s \cos (\varphi) \\ s \sin (\varphi) \\ \omega
\end{bmatrix}, \quad \xi (0) = \xi_0,
\label{eq:sys}
\end{equation}
with $\xi \inset{X} \subseteq  \Rpown{3}$, and $u = [s ~~ \omega]^\Tr \inset{PC(U)} \subset \Rpown{2}$.
We use $\mathcal{PC(U)}$ to denote that the inputs 
are piece-wise continuous and take values from a compact set $\mathcal{U}$.

\subsection{The State-Space Path-Following Problem}

We recall the path-following problem in the state space of the robot model~\eqref{eq:sys} as introduced by~\cite{faulwasser2009nonlinear}.
The path-following problem
aims at making the system \eqref{eq:sys} follow a geometric reference without
explicit timing requirements, i.e., \emph{when to be where} on the path is not specified.
The reference is given by
\begin{equation*}
    \mathcal{P} = \{\xi \inRpow{3} ~ | ~ \exists ~ \theta \inRpow{} \mapsto \xi = p(\theta)\}.
\end{equation*}
The variable $\theta(t) \inRpow{}$ is the path parameter, and $p(\theta(t)) \inRpow{3}$ is
a parameterization of $\mathcal{P}$.
Note that although $\theta$ is dependent on time, its time evolution $t \mapsto \theta$
is not specified. Thus, the control inputs $u \inset{PC(U)}$
and the timing $\theta: \Rpown{+}_0 \to  \Rpown{+}_0$ are chosen 
such that they follow the path as closely as possible.

\emph{Problem 1. (State-space path following with speed assignment)}
\begin{enumerate}
    \item Convergence to the path: the robot's state $\xi$ converges to the path $\mathcal{P}$
    such that 
    \begin{equation*}
        \lim_{t\to \infty} \|\xi(t)-p(\theta)\| = 0.
    \end{equation*}
    \item Constraint satisfaction: the constraints on the states $\xi \inset{X}$ and inputs $u \inset{U}$ are satisfied at all times. 
    \item Velocity convergence: the path velocity $\dot \theta$ converges to a predefined profile such that 
    \begin{equation*}
        \lim_{t\to \infty} \|\dot \theta(t)- v_r(t)\| = 0.
    \end{equation*}
\end{enumerate}
Here we consider path parametrizations of the form
\begin{equation}
   p(\theta) = [p_x(\theta) ~~ p_y(\theta) ~~ p_\varphi(\theta)]^\Tr,
    \label{eq:pathgral}
\end{equation}
\begin{equation*}
   p_\varphi(\theta) = \arctan \left(\frac{p_y^\prime}{p_x^\prime} \right), ~
p_x^\prime={\frac{\partial p_x}{\partial \theta}}, ~
p_y^\prime={\frac{\partial p_y}{\partial \theta}},
\end{equation*}
where $p_x(\theta)$ and $p_y(\theta)$ are at least twice 
continuously differentiable (see \cite{mehrez2017predictive}). 
We denote $p_{xy} = [p_x ~ p_y]^\Tr$ as the vector of Cartesian coordinates
of the path.

The path parameter $\theta$ is considered a virtual state, which is 
controlled by the virtual input $v$. Here the dynamics of $\theta$ are chosen  as a 
single integrator:
\begin{equation*} \label{eq:timing}
    \dot \theta = v, ~~ \theta(0) = \theta_0,
\end{equation*}
where $v \inset{PC(V)}$, $\mathcal{V} \doteq [0, \bar v]$, and $\bar v \inRpow{}$.

The path following problem is formulated using the augmented system 
\begin{equation*}
\dot z = f(z, w) = 
    \begin{bmatrix}
        \dot q_x \\
        \dot q_y \\
        \dot \varphi \\
        \dot \theta
    \end{bmatrix} = 
    \begin{bmatrix}
        s \cos (\varphi) \\ s \sin (\varphi) \\ \omega \\ v
    \end{bmatrix},
\end{equation*}
with the augmented state vector 
$z = [\xi^\Tr ~ \theta]^\Tr = [q_x ~ q_y ~ \varphi ~ \theta ]^\Tr \inset{Z}=\mathcal{X}\times \Rpown{+}_0$ and the augmented
input vector $w = [u^\Tr ~ v]^\Tr = [s ~ \omega ~ v]^\Tr \inset{PC(U \times V)} \subset \Rpown{3}$.

System \eqref{eq:sys} is \emph{differentially flat}, and $[q_x ~ q_y]^\Tr$ is 
one of its flat outputs~\cite{Martin97}. 
Therefore there is an input $u_r = [s_r ~ \omega_r]$ which guarantees 
that path \eqref{eq:pathgral} is followed by the system. 

The vector $u_r$ is used as a reference for the input vectors and 
can be built by observing that the first two equations of system \eqref{eq:sys}
satisfy $s^2 = \dot q_x^2 + \dot q_y^2$, and thus:
\begin{equation}
\begin{aligned}
    s_r(\theta, v) &= 
    \sqrt{\left( \totalderiv{p_x(\theta(t))}{t}\right)^2 + 
    \left( \totalderiv{p_y(\theta(t))}{t}\right)^2} \\
    &= 
    v \sqrt{\left(p_x^\prime\right)^2 + \left(p_y^\prime\right)^2}.
\end{aligned}
\label{eq:sref}
\end{equation}
 Furthermore, from the last equation of system \eqref{eq:sys}
we have $\omega = \dot \varphi$, which yields
\begin{equation}
\begin{aligned}
    \omega_r(\theta, v) =& \  
    \totalderiv{p_\varphi(\theta(t))}{t} \\
    =& \ v \left(\left(p_x^\prime\right)^2 + \left(p_y^\prime\right)^2 \right)^{-1} 
    \left(p_x^\prime p_y^{\prime \prime} - p_y^\prime p_x^{\prime \prime} \right), \\
&\textnormal{with } ~ p_x^{\prime \prime} = \partderiv{^2p_x}{\theta^2}, ~ \textnormal{ and }  ~
p_y^{\prime \prime} = \partderiv{^2p_y}{\theta^2}.
\end{aligned}
\label{eq:wref}
\end{equation}
Further details on the derivation can be found in \cite{ifat:faulwasser11a,mehrez2017predictive}.

\subsection{Model Predictive Path Following Control (MPFC)}

This section is based on the state-space MPFC scheme proposed in
\cite{faulwasser2009nonlinear}. 
For paths defined in output spaces, we refer to \cite{faulwasser2015nonlinear,faulwasser2016implementation}.

The sampling period is $\delta > 0$, and the prediction horizon is $T = N \delta$, 
with $N \in \mathbb{N}$.
The extended state at the current sampling time $t_k = k \delta$
is denoted $z_k =\begin{bmatrix}
\xi(t_k) & \theta(t_k)
\end{bmatrix}$ and the extended control input is
$w =\begin{bmatrix}
u & v
\end{bmatrix}$.
We consider the stage cost
\begin{equation*}
    \ell(z, w) = \left\| \begin{matrix}
        \xi- p(\theta) \\
        \theta
    \end{matrix} \right\|^2_Q + 
    \left\| \begin{matrix}
        u - u_r(\theta, v) \\
        v - v_r
    \end{matrix} \right\|^2_R,
\end{equation*}
with $Q=Q^\Tr \succeq 0$ and $R=R^\Tr \succ 0$, i.e., symmetric positive (semi)definite 
diagonal matrices. 
The Optimal Control Problem (OCP) to be solved repeatedly
at each sampling instant $t_k$ and using $z_k$ as parametric data reads
\begin{equation}
\begin{aligned}
  \vb{w}^* = \underset{w \inset{PC(W)}}{\text{arg min}}
  & ~~ \int_{0}^{T} \ell({z} (\tau), {w}(\tau)) \textrm{d}\tau 
  \\
  \text{subject to}
  & ~~ \dot z(\tau) = f(z(\tau),w(\tau)), \quad z(0) = z_k, \\
  & ~~ z(\tau) \inset{Z}, ~ w(\tau) \inset{W}. \\
  \end{aligned}
    \label{eqn:ocp}
\end{equation}
Although this OCP is formulated in continuous time, 
our MPFC \emph{implementation} is done in discrete time
with $\vb{w}^* = \{w_0^*, \ldots, w_{N-1}^*\} \inset{W}^N$ a
sequence of $N$ input vectors.
Typically, in MPC we only apply to the controlled system the first vector 
$w=w_0^*$ in the sequence $\vb{w}^*$. 
The MPFC feedback controller based on \eqref{eqn:ocp} can be expressed as the function
\begin{equation}
    w = \begin{bmatrix}
    u \\ v
\end{bmatrix}  =   \mathbb{M}(z).
    \label{eqn:mpcmap}
\end{equation}
Observe that $w$ entails the robot command $u$ and the virtual control $v$, which controls the evolution of the path parameter $\theta$, cf. \eqref{eq:timing}. Hence only $u$ is applied to the robot. 

\section{Feedforward Neural Networks}
\label{sec:qdnn}

Next, we recall the basics of how a function
can be approximated by feedforward neural networks, 
the advantages of using deep architectures, and 
how to quantize them.

\subsection{Deep Neural Networks}
The use of feedforward Neural Networks (NN)
is motivated by their universal function approximation properties \cite{leshno1993multilayer}. 
In particular, we are interested in approximating the MPFC feedback \eqref{eqn:mpcmap}.
Our goal is to train an NN that approximates $\mathbb{M}(z)$ 
by defining the mapping $w^D=\mathbb{D}(z; \Theta)$,
where $\Theta$ represents a set of $N_\Theta$ unknown parameters, which are learned during \emph{training}.
Once we have a trained network, we can use the $\mathbb{D}(z; \Theta)$ to \emph{infer} 
the values of $w^D\approx \mathbb{M}(z)$.

To train our network, we rely on a training data set
\begin{equation*}
\mathcal{T} = \{\nu^1, \nu^2, \ldots, \nu^{N_T}\}, ~ \textnormal{with }
\nu^j = \begin{bmatrix}
    z^j \\ \mathbb{M}(z^j)
\end{bmatrix} \inRpow{7},
\end{equation*}
$j \inset{J} = \{1, \ldots, N_T\}$, and $N_T$ is large enough.
The training algorithm aims to find the values of 
$\Theta$ that make $\mathbb{D}(z^j; \Theta) \approx \mathbb{M}(z^j), \forall ~ j \inset{J}$
using some statistical measure like the Mean Squared Error (MSE).
It is common to use a gradient-based optimization algorithm during training to minimize the MSE.
The trained network is said to \emph{generalize} well if $\mathbb{D}(z; \Theta)$ 
is still a good approximation of $\mathbb{M}(z)$
for values of $z$ not seen during training,
in particular those 
relevant to the application.

In general, an NN consists of $H+2$ layers: one input layer, one output layer,
and $H \geq 1$ hidden layers. Each layer $k$ consists of $n_k$ units called neurons. 
Commonly, if there are only one or two hidden layers, the network is referred to as \emph{shallow},
otherwise, it is called a \emph{Deep} Neural Network (DNN).
The advantage of a DNN, compared to a shallow network, is that it can
approximate a function like \eqref{eqn:mpcmap} with similar accuracy 
but with fewer parameters $N_\Theta$ as fewer neurons (and hence parameters) are considered per layer. We refer to \cite{goodfellow2016deep} for details.

Starting with the input $z = h^0$ as the first layer,
the output of layer $k=1,2,\ldots, H+1$ is
\begin{equation}
    h^k = \beta(b^k +  W^k h^{k-1}),
    \label{eq:NNLayer}
\end{equation}
with $b^k \inRpow{n_k}$ a vector called \emph{bias} and $W^k \inRpownm{n_k}{n_{k-1}}$
a matrix called \emph{weights}, and the function $\beta(\cdot)$ is a saturating 
\emph{activation} function.
The last layer is the output layer $w^D=h^{H+1}$.
Note that $\Theta = \{b^1, W^1, \ldots, b^{H+1}, W^{H+1}\}$, and 
the number of parameters of the network is given by:
\begin{equation*}
N_\Theta = \sum_{k=1}^{H+1} n_k(1 + n_{k-1}).
\end{equation*}

For example, a network with $1$ hidden layer would be
described as
$
w^D = \mathbb{D}(z; \Theta) = \beta(b^2 + W^2 \beta(b^1 + W^1 z)).
$

A frequently used activation function is the  
Rectifying Linear Unit (ReLU) (\cite[p. 171]{goodfellow2016deep}), defined as
$\beta(h) = \max(\vb{0}, h)$, where $\max(\cdot)$ is computed element-wise.
Other common activation functions include the tangent hyperbolic and the sigmoid function. 


\subsection{Network Training}

In practice, to find the set of  parameters $\Theta$ that make $\mathbb{D}(\cdot)$ approximate 
$\mathbb{M}(\cdot)$ sufficiently well
the higher-level set of so-called \emph{hyper-parameters}  needs to be determined. 
Common hyper-parameters include the network architecture ($H$, $n_k$, $\beta$),
and the gradient-based optimization algorithm parameters (e.g., the step size, also called the \emph{learning rate}) 
to name just a few.
A suitable combination of hyper-parameters is typically determined experimentally \cite{bergstra2012random}.

It is helpful to normalize the training set to improve the numerical properties of 
the network. 
Here, we represent the training set as a matrix $\mathcal{T} \inRpownm{7}{N_T}$ 
for simplicity in notation.
For each column $j$, and row $i$ of $\mathcal{T}$ we have: 
\begin{equation*}
    \bar \nu^j_i = N(\nu^j_i; \mu_{i}, \sigma_{i}) = \frac{\nu^j_i-\mu_{i}}{\sigma_{i}},
\end{equation*}
where $\mu_i$ is the mean and $\sigma_i$ is the standard deviation 
of row $i$. Note that $\nu^j$ represents column $j$ of $\mathcal{T}$.
After applying this transformation, we obtain a normalized 
data set $\mathcal{T}_N$ that has each row $i$ with $\bar \mu_i=0$ and 
$\bar \sigma_i=1$. To recover the original set $\mathcal{T}$, we apply the inverse transformation:
\begin{equation*}
    \nu_i^j = N^{-1}(\bar \nu_i^j; \mu_{i}, \sigma_{i}) = \bar \nu_i^j \sigma_{i} + \mu_{i}.
\end{equation*}

These operations must be applied to the extended robot state $z = \begin{bmatrix} \xi & \theta\end{bmatrix} \inRpow{4}$ 
and the extended input vector 
$w^D = \begin{bmatrix} u & v\end{bmatrix}\inRpow{3}$ during inference.
That is $\bar z_i = N(z_i; \mu_{i}, \sigma_{i})$, for $i=1,2,3,4$,
and $w^D_i = N^{-1}(\bar w_i; \mu_{i+4}, \sigma_{i+4})$, for $i=1,2,3$
(refer to Fig. \ref{fig:dnn}).

\subsection{Quantized DNN (QDNN)}

Quantization refers to storing the parameters
of the network (weights and biases) as integer values.
The main advantages are reduced  memory required to store the $N_\Theta$ parameters,
faster execution, and higher energy efficiency during inference.
The main disadvantage is the loss of accuracy in the inference \cite{gholami2021survey}.

It is common to use an $8$-bit integer representation (i8) to store the parameters set $\Theta$. 
The network is trained first using floating point numbers often with single precision (32 bits).
After the training is completed, 
the parameters $\Theta$ are \emph{quantized} to an i8 approximation.
There are different quantization methods \cite{gholami2021survey}. Here we have
used a uniform asymmetric quantization.
That means that during inference, the \emph{normalized} inputs $\bar z$ 
in the network must be 
transformed from a floating point number to an integer using 
\begin{equation}
 \hat z = Q(z; \tilde a, \hat b) = \mathrm{i8}(\tilde a \bar z)  + \hat b,
 \label{eq:NNQuant}
\end{equation}
where $\tilde a$ is a floating point scaling, $\hat b$ is an integer offset,
and $\mathrm{i8}$ refers to a mapping from floating point to 8-bit integer 
representation.
Similarly, 
the output of the network $\hat w$ must be transformed from an 8-bit integer 
to a floating-point 
normalized output $\bar w$, i.e., it must be \emph{dequantized} using
\begin{equation}
 \bar w = Q^{-1}(\hat w; \tilde c, \hat d) = \mathrm{f32}(\hat w - \hat d) \tilde c,
 \label{eq:NNDequant}
\end{equation}
where $\tilde c$ is a floating point scaling, $\hat d$ is an integer offset,
and $\mathrm{f32}$ refers to a mapping from an 8-bit integer to a single-precision 
floating-point 
representation.
The scaling and offset parameters are determined
during the quantization of $\Theta$.
Fig. \ref{fig:dnn} depicts how the robot state $z$ (input to the network)
and input vector $w^D$ (output of the network) are numerically transformed.

\section{QDNN-based MPFC}
\label{sec:qdnn-mpfc}

We now turn to a practical approach to
approximate the MPFC problem presented in Section \ref{sec:mpfc_robot}
using QDNNs as described in Section \ref{sec:qdnn}.
We denote this approach as MPFC-QDNN.
This section also discusses how to augment the MPFC-QDNN 
with an online feedback controller 
to improve the accuracy of the path-following control.
We denote this approach as MPFC-QDNN+P.

\subsection{Generating a Training Set for MPFC}

Although it is possible to find a network
that approximates $\mathbb{M}(z) ~ \forall ~ z \inset{Z}$, 
this typically would require a network and set $\mathcal{T}$ larger than necessary 
for the path-following problem.
Under normal circumstances, a mobile robot following a path will mostly take poses $\xi$ 
that are close to the reference path $p(\theta)$. 
Based on this, a smaller set $\mathcal{Z}_C \subset \mathcal{Z}$
can be used to significantly reduce the size of the network and the training set, 
without affecting the performance of the MPFC near the path.
However, if the robot is driven far away from the path (e.g., due to large disturbances),
the MPFC-QDNN may not be able to bring the robot back to following the path.

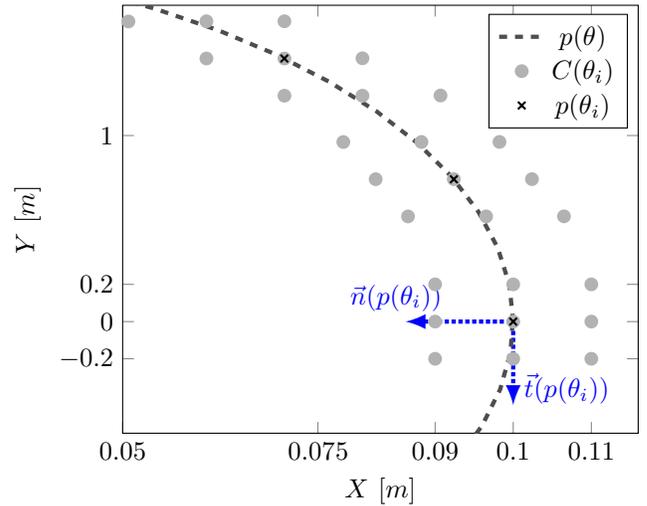
\begin{figure}[t]
\centering
\pgfplotstableread{
    x y
    0.1 0
    0.0924 0.765
    0.0707 1.414 
}\pxpy

\begin{tikzpicture}
    \begin{axis}[xmin=0.05, ymin=-0.60, ymax=1.7, 
        xticklabel style={
              /pgf/number format/precision=3,
              /pgf/number format/fixed}, 
        xtick={0.11, 0.1, 0.09, 0.075, 0.05}, 
        ytick={-0.2, 0, 0.2, 1},
        xlabel={$X ~ [m]$},
        ylabel={$Y ~ [m]$},
        ]
        \node (source) at (axis cs:0.1, 0){};
        \node (Np) at (axis cs:0.085, 0){};
        \node (Tp) at (axis cs:0.1, -.5){};
        \draw[-latex, densely dotted, ultra thick, blue](source)--(Np)
        node[above] {$\vec{n}(p(\theta_i))$};
        \draw[-latex, densely dotted, ultra thick, blue](source)--(Tp)
        node[above right] {$\vec{t}(p(\theta_i))$};
        \addplot [black!70, ultra thick, dashed, no marks] table [x=pos-x, y=pos-y, col sep=comma] {figs/pathparam.csv};
        \addlegendentry{$p(\theta)$}
        \addplot [black!30, very thick, only marks] table [x=pos-x, y=pos-y, col sep=comma] {figs/corridor.csv};
        \addlegendentry{$C(\theta_i)$}
        \addplot [black, thick, only marks, mark=x] table{\pxpy};
        \addlegendentry{$p(\theta_i)$}
    \end{axis}
\end{tikzpicture}
\caption{Simplified 2-dimensional visualization of data point used to build the 
training set $\mathcal{T}$.
The vectors $\vec{n}$, $\vec{t}$, $\vec{o}$ (coming out of the page) are orthonormal.
Only $p_x$ and $p_y$ are shown 
(the orientation $p_\varphi$ is not depicted).
The figure shows the corridor for three values of $\theta_i$, and at each point
$p(\theta_i)$ (cross) a corridor $C(\theta_i)$ of $9$ points (dots) is constructed.
The width ($0.02$ m in the figure),
and the length ($0.4$ m) of the corridor are measured normal (along $\vec{n}$) 
and tangential (along $\vec{t}$) to the path
at $p(\theta_i)$, respectively.
}
\label{fig:corridor}
\end{figure}

To generate a set $\mathcal{T}$ appropriate for MPFC, 
we propose to use a corridor centered around the path
(see Fig. \ref{fig:corridor}). 
To build the set $\mathcal{T}$, we select specific values of the path parameter 
$\theta_i$, $i=0,1,...,N_p$, and compute the
path vector $p(\theta_i)$. At each $\theta_i$, 
we build a corridor $C(\theta_i) \inRpow{3 \times N_c}$ 
using a set of $N_c$ points in the vicinity of
$p(\theta_i)$. 

We propose a corridor in the form of a cuboid centered
around $p(\theta_i)$ along the orthonormal vectors $\vec{t}, \vec{n}, \vec{o}$ (see Fig. \ref{fig:corridor}),
with width $2c_W$, length $2c_L$, and height $2c_H$.
The points $C^j(\theta) = [p^j_t ~ p^j_n ~ p^j_o]^\Tr$
are equidistant along each axis, with 
\begin{equation*}
    \begin{bmatrix}
        -c_W\vec{t} ~ \\ - c_L\vec{n} \\ - c_H \vec{o}
    \end{bmatrix}
    \leq
    \begin{bmatrix}
        p^j_t \\ p^j_n \\ p^j_o
    \end{bmatrix}
    \leq
    \begin{bmatrix}
        c_W\vec{t} ~ \\ c_L\vec{n} \\  c_H \vec{o}
    \end{bmatrix}.
\end{equation*} 

The set $\mathcal{Z}_C$ has $N_p N_c$ elements.
The corridor can be defined in many different ways (e.g., using randomly selected points
inside an ellipsoid). 
Here we have presented one way 
that has worked well in our experiments (a cuboid grid with equidistant points). Determining the optimal way to construct the
set $\mathcal{Z}_C$ is beyond the scope of this work.

The size of the corridor plays an important role in how
well the MPFC-QDNN can follow the path in practice. 
If the corridor is too narrow, the network is not able to follow the path at all, due
to inevitable errors inherent in any feedback control system.
A broad corridor is thus preferred. However, that may require more data points 
in the set, and perhaps a larger network, 
to make the approximation $\mathbb{D}(\cdot)$ useful.

\subsection{Path-Following Error}

\begin{figure}[t]
\centering
\pgfplotstableread{
    x y
    0. 2.
}\pxpy

\pgfplotstableread{
    x y
    0.03 1.5
}\qxqy

\begin{tikzpicture}
    \begin{axis}[xticklabel style={
              /pgf/number format/precision=4,
              /pgf/number format/fixed
              }, 
              xmin=-0.02, xmax=0.07, ymin=0.9, ymax=2.2, 
        xtick={-0.05, 0.0, 0.05}, 
        ytick={1, 2},
        xlabel={$X$ [m]},
        ylabel={$Y$ [m]},
        legend pos=south east,
        ]
        \node (source) at (axis cs:0., 2.){};
        \node (Np) at (axis cs:0, 1){};
        \node (Tp) at (axis cs:0.05, 2){};
        \node (projn) at (axis cs:0, 1.5){};
        \node (projt) at (axis cs:0.03, 2){};
        \node (robot) at (axis cs:0.03, 1.5){};
        \node (mind) at (axis cs:0.066, 1.502){};
        \draw[-latex, densely dotted, ultra thick, blue](source)--(Np)
        node[above right] {$\vec{n}(p_\varphi)$};
        \draw[-latex, densely dotted, ultra thick, blue](source)--(Tp)
        node[below] {$\vec{t}(p_\varphi)$};
        \draw[-latex, densely dotted, ultra thick, red](source)--(robot)
        node[above right] {$e(q,p)$};
        \draw[dashed](robot)--(projt);
        \draw[-latex,ultra thick](source)--(projt)
        node[above left] {$e_t(q,p)$};
        \draw[dashed](robot)--(projn);
        \draw[-latex,ultra thick](source)--(projn)
        node[above left] {$e_n(q,p)$};
        \addplot [black!40, ultra thick, loosely dashed, no marks] table [x=pos-x, y=pos-y, col sep=comma] {figs/pathparam.csv};
        \addlegendentry{$p(\theta(t))$}
        \addplot [black, thick, only marks, mark=x] table{\pxpy};
        \addlegendentry{$p_{xy}(\theta_k)$}
        \addplot [black, thick, only marks, mark=star] table{\qxqy};
        \addlegendentry{$q(t_k)$}
    \end{axis}
\end{tikzpicture}
\caption{
At any given time $t_k$, the robot's position in the global frame $XY$ is given by $q(t_k)$. 
The unit vectors $\vec{t}(\theta_k)$ (not shown at scale), and
$\vec{n}(\theta_k)$ are tangential and normal to the path at point $p_{xy}(\theta_k)$, respectively, 
and $\theta_k = \theta(t_k)$. 
The vector $e(q,p)=q - p_{xy} = e_t(q,p) \vec{t}(p_\varphi) + e_n(p,q) \vec{n}(p_\varphi)$ 
is the current robot's position with respect to the path.
}
\label{fig:projection}
\end{figure}
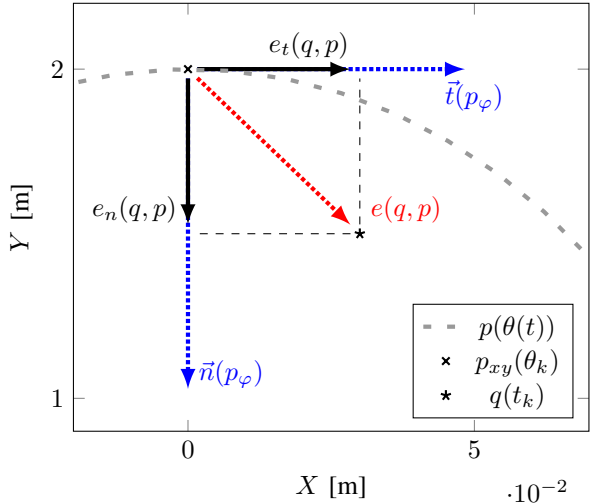

Although MPFC can follow the reference path $\mathcal{P}$ very accurately, 
at any time $t$ there might be an error $e$ in the robot's Cartesian position $q(t)$ with respect 
to the reference point in the path $p(\theta(t))$. In the $XY$ coordinates, the error is given by:
\begin{equation*}
    e_{XY}(q(t),p_{xy}(\theta(t)))= q(t) - p_{xy}(\theta(t)).
\end{equation*}
This error vector can be expressed in the basis formed by
the orthonormal vectors 
$\vec{t}(p_\varphi(\theta))$ and $\vec{n}(p_\varphi(\theta))$, which are tangential and normal to the path
at $p_{xy}(\theta)$, respectively (refer to Fig. \ref{fig:projection}).
That is:
\begin{equation}
e(q(t), p(\theta(t))) = e_n(q,p) \vec{n}(p_\varphi) + e_t(q,p) \vec{t}(p_\varphi),
    \label{eq:errorvec}
\end{equation}
where the scalars $e_n$ and $e_t$ are the projection of $e_{XY}$
onto each orthonormal vector, computed by the dot product
\begin{equation*}
    \begin{aligned}
    e_n(q,p) = \left(q - p_{xy}\right)^\Tr \vec{n}(p_\varphi), \\
    e_t(q,p) = \left(q - p_{xy}\right)^\Tr \vec{t}(p_\varphi).
    \end{aligned}
\end{equation*}
In the case of an ellipse, the tangential and normal
vectors are given by:
\begin{equation*}
   \vec{t}(p_\varphi) = \begin{bmatrix}
        \cos (p_\varphi) \\ \sin (p_\varphi)
    \end{bmatrix}, ~~
    \vec{n}(p_\varphi) = \begin{bmatrix}
        -\sin (p_\varphi) \\ \cos (p_\varphi)
    \end{bmatrix}.
\end{equation*}

\subsection{Augmented Control Scheme}

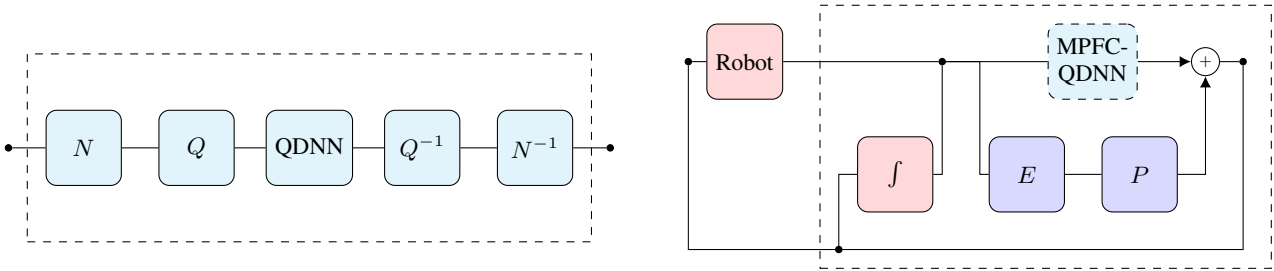
\begin{figure*}
    \begin{subfigure}[b]{0.5\textwidth}
    \centering
\tikzset{%
  >={Latex[width=1.5mm,length=1.5mm]},
            base/.style = {rectangle, rounded corners, draw=black,
                           minimum width=1cm, minimum height=1cm,
                           fill=cyan!10,
                           text centered, font=\sffamily, font=\small},
         xyz/.style = {base, minimum width=1cm, fill=blue!15,
                           font=\ttfamily, font=\small},
         sys/.style = {xyz, fill=red!15},
         txt/.style = {opacity=0, text opacity=1, font=\small},
         crc/.style = {circle, fill=black, draw=black, inner sep=0.03cm},
}

\begin{tikzpicture}[node distance=0.5cm,
    every node/.style={fill=white, font=\sffamily, font=\tiny}, align=center]
  \node (x0)    [circle, fill=black, draw=black, inner sep=0.03cm] {};
  \node (r) [rectangle, dashed, opacity=1, fill=none, draw=black,
     minimum width=7.5cm, minimum height=2.5cm, right of=x0,
     xshift=3.5cm, yshift=-0.0cm] {} node [txt, above right, xshift=0.5cm, yshift=0.7cm] {MPFC-QDNN};
  \node (x)   [txt, right of=x0, above, xshift=-0.5cm, above] {$z$};
  \node (nrm)   [base, right of=x0, xshift=0.5cm] {$N$};
  \node (qnt)   [base, right of=nrm, xshift=1.0cm] {$Q$};
  \node (dnn)   [base, right of=qnt, xshift=1.cm] {QDNN};
  \node (den)   [base, right of=dnn, xshift=1.cm] {$Q^{-1}$};
  \node (deq)   [base, right of=den, xshift=1.0cm] {$N^{-1}$};
  \node (uf) [crc, right of=deq, xshift=0.5cm] {}; 
  \node (u) [txt, right of=uf, above, xshift=-0.4cm] {$w^D$};
  \draw[-] (x0) -- (nrm);
  \draw[-] (nrm) -- node[txt, above] {$\bar z$}(qnt);
  \draw[-] (qnt) --  node[txt,above] {$\hat z$} (dnn);
  \draw[-] (dnn) -- node[txt,above] {$\hat w$} (den);
  \draw[-] (den) -- node[txt, above] {$\bar w$}(deq);
  \draw[-] (deq) -- (uf);
  
  \end{tikzpicture}
\hfill
\caption{The quantized deep neural network inference chain 
used to approximate the MPFC (denoted MPFC-QDNN on the right figure).}
\label{fig:dnn}
    \end{subfigure}
    \begin{subfigure}[b]{0.44\textwidth}
    \centering
\tikzset{%
  >={Latex[width=1.5mm,length=1.5mm]},
            base/.style = {rectangle, rounded corners, draw=black,
                           minimum width=1cm, minimum height=1cm,
                           text centered, font=\sffamily, font=\small},
         xyz/.style = {base, minimum width=1cm, fill=blue!15,
                           font=\ttfamily, font=\small},
         sys/.style = {xyz, fill=red!15},
         dnn/.style = {xyz, fill=cyan!10},
         txt/.style = {opacity=0, text opacity=1, font=\small},
         crc/.style = {circle, fill=black, draw=black, inner sep=0.03cm},
}

\begin{tikzpicture}[node distance=0.5cm,
    every node/.style={fill=white, font=\sffamily, font=\tiny}, align=center]
  \node (x0)    [crc] {};
  \node (x1)    [crc, below of=x0, xshift=2cm, yshift=-2cm] {};
  \node (rbt) [sys, right of=x0, xshift=0.25cm] {Robot};
  \node (path) [sys, right of=x1, xshift=0.25cm, yshift=1.cm]  {$\int$};
  \node (xp)    [crc, right of=rbt, xshift=2.125cm] {};
  \node (dnn)   [dnn, dashed, right of=xp, xshift=1.5cm] {MPFC- \\ QDNN};
  \node (up)    [circle, fill, draw=black, inner sep=0.05cm, above, right of=dnn, xshift=1.cm] {$+$};
  \node (err)   [xyz, right of=path, xshift=1.25cm] {$E$};
  \node (pid)   [xyz, right of=err, xshift=1.cm] {$P$};
  \node (u) [txt, right of=up, above, xshift=0.cm] {$w$};
  \node (uf) [crc, right of=up, xshift=0.cm] {}; 
  \node () [txt, left of=rbt, above, xshift=-0.25cm] {$s$\\$\omega$}; 
  \node () [txt, left of=path, above left, xshift=0.cm]  {$v$};
  \node () [txt, right of=rbt, above, xshift=0.25cm] {$\xi$}; 
  \node () [txt, right of=path, below right]  {$\theta$};
  \draw[-] (xp) |-  node[txt,below right, xshift=.5cm] {$z$} (dnn);
  \draw[-] (err) -- node[txt,above] {$e$} (pid);
  \draw[->] (pid) -|  node[txt,below] {$w^P$} (up);
  \draw[->] (dnn) -- node[txt,above, xshift=-0.cm] {$w^D$} (up);
  \draw[-] (up) -- (uf);
  \draw[-] (x0) -- (rbt);
  \draw[-] (x1) |- (path);
  \draw[-] (x1) -| (x0);
  \draw[-] (rbt) -- (xp);
  \draw[-] (path) -| (xp);
  \draw[-] (uf) |- ++ (-2.5, -2.5) -| (x1);
  \draw[-] (xp) -- ++ (.5, 0) |- (err);

  \node (r) [rectangle, dashed, opacity=1, fill=none, draw=black,
     minimum width=6.cm, minimum height=3.5cm, left of=x1,
     xshift=3.25cm, yshift=1.5cm] {} 
     node [txt, above right of=x1, xshift=0.7cm, yshift=2.6cm] {MPFC-QDNN+P};
  
  \end{tikzpicture}
\label{fig:dnnp}
\caption{The proposed controller approach (MPFC-QDNN+P)}
    \end{subfigure}
\caption{Block diagram of the proposed combined controller.
Inside the dashed block (MPFC-QDNN) is the MPFC controller approximated by a QDNN.
The extended state $z$ is fed into the MPFC-QDNN. The state is first normalized  
($\bar z$) then quantized ($\hat z$), and finally fed into the QDNN block for inference
of the approximate control input $\hat w$. Dequantization and denormalization 
are applied to get the approximated augmented control input $w^D$.
The block $E$ computes the robot's path error vector $e$ \eqref{eq:errorvec}
used by the P controllers. 
The vector $w^P$
is added to $w^D$ to compute the input $w$.
The forward velocity $s$, and angular velocity $\omega$ 
are applied to the robot, whereas
the path velocity $v$ is integrated 
to compute the path variable $\theta$.}
\label{fig:ctl}
\end{figure*}

Due to the MPFC-QDNN being an approximation of the MPFC,
the path-following error $e$ resulting from $\mathbb{D}(z; \Theta)$ 
is significantly larger than the 
error observed under the original MPFC controller $\mathbb{M}(z)$
(see Section \ref{sec:results}).
To compensate this error
we extend the MPFC-QDNN controller with an additional linear feedback which acts on the tangential
component $e_t$ through the forward speed of the robot $s$, 
and on the 
normal component $e_n$ through the robot's angular speed $\omega$. 
Put differently,  the compensation term $w^P$
is added to the control vector, i.e., $w = w^D + w^P$.
Here $w^P = [s^P ~ \omega^P ~ 0]^\Tr$, with $s^P = P_t e_t$, and 
$\omega^P  = P_n e_n$, 
where $P_t$, and $P_n$ are the proportional gains
(see Fig. \ref{fig:ctl}).
We denote this approach MPFC-QDNN-P.
We selected
static feedback mainly due to its simplicity and effectiveness as shown in 
Section \ref{sec:results}.

\subsection{Implementation}
\label{sec:impl}

We consider an ellipse as the path (see Fig. \ref{fig:diffdrvrbt}), 
which is defined by the parametrization 
\begin{equation*}
    p(\theta) = \begin{bmatrix}
        0.1 \cos(\theta) &
        2 \sin(\theta) &
        \arctan \left(\frac{-0.1 \sin (\theta)}{2 \cos (\theta)} \right)
    \end{bmatrix}^\Tr,
\end{equation*}
which yields the input references \eqref{eq:sref} and \eqref{eq:wref} as
\begin{align*}
    s_r (\theta, v) &= 2v \sqrt{1- 0.9975 \sin^2(\theta)}, \\
    \omega_r (\theta, v) &= 2 v \left(40 - 39.9 \sin^2(\theta)\right)^{-1}.
\end{align*}

To generate the training set $\mathcal{T}$, we use a corridor consisting of a cuboid of 
width $2c_W=0.02$, length $2c_L=0.2$, and height $2c_H=\frac{2\pi}{3}$. Each axis is split
into $5$, $5$, and $40$  equidistant points ($N_c=1000$), respectively.    
We split the path in $N_p=4000$ equidistant segments between $0\leq \theta \leq 2\pi$,
which corresponds to a full turn around the path.
The subset of states in the corridor $\mathcal{Z}_C$ consists of $N_p N_c = 4E6$ points.

To solve the MPFC problem \eqref{eqn:ocp}, and consequently
find $w = \mathbb{M}(z), ~ \forall ~ z \inset{Z}_C$ according to
$\eqref{eqn:mpcmap}$, we use the
Optimization Engine (OpEn) 
\cite{open2020}, a fast solver for optimal control problems. 
The training set $\mathcal{T}$ consists of $N_p N_c$ 
pair of vectors $z$, $\mathbb{M}(z)$ for all $z \inset{Z}_C$.
We use a discretization time $\delta = 0.01$ s, and a horizon length $T=0.6$ s
in \eqref{eqn:ocp}.

We use a random search approach to find the hyper-parameters of a network 
that is a good approximation to $\mathbb{M}$ under the constraint that the 
number of parameters $N_\Theta$
should remain \emph{small}. i.e.,  
to reduce the size of the network in the MCU's ROM.
Random search typically delivers better results than manual or grid search
for the same amount of computation during training 
\cite{bergstra2012random}.
The selected hyper-parameters were the number of hidden layers $H$,
the number of units in each hidden layer $n_k$, $k=1,\ldots, H$, 
and the learning rate of the optimization algorithm  
(see the Appendix).
To find the hyper-parameters we use KerasTuner \cite{omalley2019kerastuner}.
To perform the training and the quantization of the network we use 
the deep learning framework 
Keras/TensorFlow \cite{abadi2016tensorflow}, \cite{chollet2015keras}.

We implement a Hardware-In-the-Loop (HIL) simulation where 
the MPFC-QDNN+P is deployed on an STM32F407 MCU, which is based on a Cortex-M4 processor
core running at 168 MHz, which includes a single-precision floating-point unit and $1$ MB flash ROM.
The robot dynamics are simulated on a PC, see the Appendix for details.

The QDNN consists of 9 hidden layers 
with roughly 4700 parameters,
using 8-bit integers to store the parameters
(i.e. $1$ parameter requires $1$ byte of ROM).
The quantized parameter set $\Theta$ requires less than $5$ kB 
of the MCU's flash memory. 
\section{Results}
\label{sec:results}

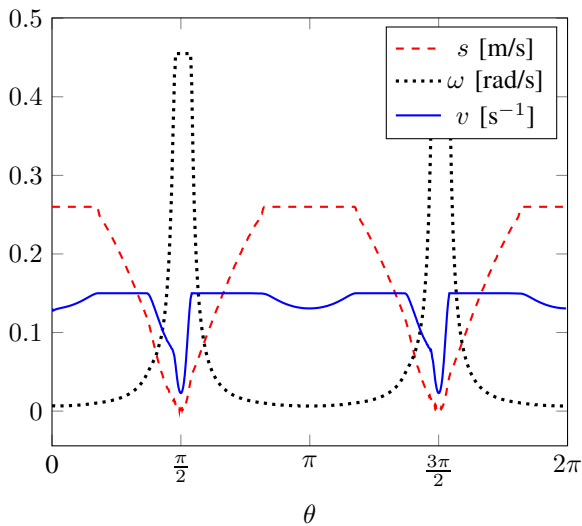
\begin{figure}[t]
\centering
\begin{tikzpicture}
    \begin{axis}
        [xlabel = $\theta$,
        xmin=0, xmax=6.28,
        xtick={0, 1.57, 3.14, 4.71, 6.28}, 
        xticklabels={0, $\frac{\pi}{2}$, $\pi$, $\frac{3\pi}{2}$, $2\pi$},
            skip coords between index={518}{1000}]
        \addplot [red, thick, dashed, no marks] table [x=pathparam, y=linvel, col sep=comma] {figs/NMPC_xu.csv};
        \addlegendentry{$s$ [m/s]}
        \addplot [black, very thick, dotted, no marks] table [x=pathparam, y=angvel, col sep=comma] {figs/NMPC_xu.csv};
        \addlegendentry{$\omega$ [rad/s]}
        \addplot [blue, thick, no marks] table [x=pathparam, y=pathvel, col sep=comma] {figs/NMPC_xu.csv};
        \addlegendentry{$v$ [s$^{-1}$]}
    \end{axis}
\end{tikzpicture}
\caption{Control inputs vs. path parameter. When the path curvature is
tight (near $\frac{\pi}{2}$, and $\frac{3\pi}{2}$)
the path speed $v$, and the robot's forward speed $s$ are reduced.
The path speed $v$ is also reduced when the input constraints are active (near $\pi$). 
}
\label{fig:inputs}
\end{figure}

\begin{figure}[t]
\centering
\begin{tikzpicture}
    \begin{axis}[
        xticklabel style={
              /pgf/number format/precision=3,
              /pgf/number format/fixed}, 
        xlabel={$X$ [m]},
        ylabel={$Y$ [m]},
        ylabel style={yshift=-1.0em},
            skip coords between index={6000}{12000},
        ]
        \addplot [black!70, ultra thick, loosely dashed, no marks] table [x=pos-x, y=pos-y, col sep=comma] {figs/pathparam.csv};
        \addlegendentry{OpEn}
        \addplot [blue, very thick,  dotted, no marks] table [x=pos-x, y=pos-y, col sep=comma] {figs/non_quantized_NN-S.csv};
        \addlegendentry{DNN}
        \addplot [green, very thick, dash dot, no marks] table [x=pos-x, y=pos-y, col sep=comma] {figs/quantized_NN-S.csv};
        \addlegendentry{QDNN}
        \addplot [red, thick, no marks] table [x=pos-x, y=pos-y, col sep=comma] {figs/quantized_NN-S_P.csv};
        \addlegendentry{QDNN+P}
        \node () at (axis cs:0.1, 0) [left] {$\theta=0$};
        \node () at (axis cs:0., 2) [below] {$\theta=\frac{\pi}{2}$};
        \node () at (axis cs:-0.1, 0.) [right] {$\theta=\pi$};
        \node () at (axis cs:0., -2) [above] {$\theta=\frac{3\pi}{2}$};
    \end{axis}
\end{tikzpicture}
\caption{The path followed by the robot using 
different MPFC implementations: OpEn (visually indistinguishable from the reference path $p(\theta)$), 
the approximation using 
a deep neural network (DNN, 32-bit float), 
a quantized deep neural network (QDNN, 8-bit integer), 
and the proposed QDNN plus P control approach (QDNN+P).
}
\label{fig:path}
\end{figure}
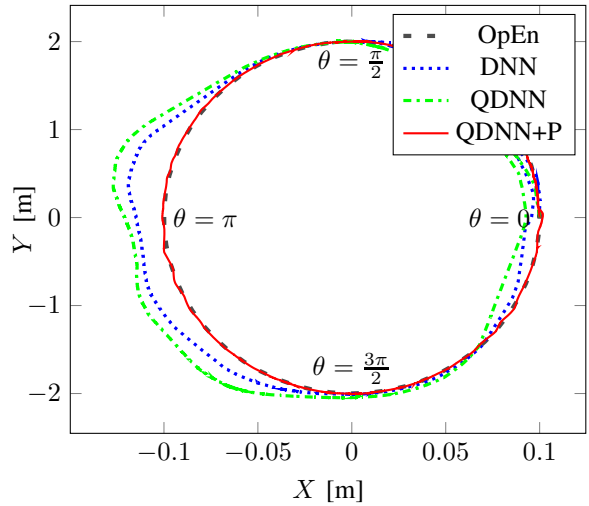

As our reference implementation (denoted MPFC-OpEn), 
we use OpEn (the same solver used for training)
to solve the MPFC problem \eqref{eqn:ocp}.
The advantages of MPFC are illustrated in Fig. \ref{fig:inputs}. The input $w$ computed by 
OpEn to steer the robot along the path in Fig. \ref{fig:path} shows that
when the path curvature is tight, i.e. top ($\theta=\frac{\pi}{2}$)
and bottom ($\theta=\frac{3\pi}{2}$) of the ellipse, 
the path speed $v$ is reduced, and  consequently the robot's forward 
speed $s$ is also reduced.  This 
allows the robot to follow the tight curve. 
Similarly, $v$ is reduced when 
the constraints on $s$ are active (e.g. $\theta=\pi$)
because the robot cannot otherwise closely follow the path.

\begin{table}
    \centering
    {\renewcommand{\arraystretch}{1.2}
\begin{tabular}[h]{|l|c|c|c|c|}
    \hline
    Implementation & Mean [s] & Std. [s] & Worst [s] \\
    \hline
    OpEn (PC) & 1.2E-3 & 5.9E-4 & 7.8E-3\\
    QDNN+P (PC) & 7.3E-6 & 2.9E-6 & 3.2E-5\\
    \hline
    OpEn (MCU) & - & - & - \\
    QDNN+P (MCU) & 2.3E-4 & 2.1E-6& 2.4E-4 \\
    \hline
\end{tabular}
    }
\caption{Mean, standard deviation (Std.), worst-case execution (Worst) time in seconds for 
10000 steps close to the path. 
The table is split into MPFC implementations on a personal computer (PC, top part) and 
a microcontroller (MCU, bottom part). Note that the execution of QDNN+P on the 
MCU is temporally deterministic. In the PC case, the QDNN+P has more variability
due to the operating system. }
\label{tbl:exec-time}
\end{table}

The main advantage of using a DNN on an MCU is that it is relatively easy to implement
quantization \eqref{eq:NNQuant}, inference \eqref{eq:NNLayer}, and 
dequantization \eqref{eq:NNDequant} sequentially for all layers in the 
network.  
Furthermore, for a small network like the one used here ($4700$ parameters),
the inference is executed much faster 
than solving the OCP \eqref{eqn:ocp}.

Table \ref{tbl:exec-time} 
shows the execution time for different implementations of
MPFC.
Our experiments ran on a PC with Ubuntu Linux 22.04-LTS,
and a x86-64 processor with a $2.4$ GHz clock.
Compared to MPFC-OpEn,
the average execution time of the MPFC-QDNN+P implementation is about three orders 
of magnitude faster on the PC.

The QDNN implementation using $8$-bit integers requires on average $230$ microseconds
to execute on the MCU.
Currently, running OpEn on an MCU is not supported. 

\begin{figure}[t]
\centering
\begin{tikzpicture}
    \begin{axis}
        [
            legend pos= south east,
            skip coords between index={518}{10000},
        xmin=0, xmax=6.28,
                    xtick={0, 1.57, 3.14, 4.71, 6.28}, 
        xticklabels={0, $\frac{\pi}{2}$, $\pi$, $\frac{3\pi}{2}$, $2\pi$},
            height=4cm, width=\columnwidth]
        \addplot [no marks] table [x=pathparam, y=err, col sep=comma] {figs/NMPC_xu.csv};
        \addlegendentry{OpEn}
    \end{axis}
\end{tikzpicture}
\begin{tikzpicture}
    \begin{axis}
        [
        legend pos= south east,
            skip coords between index={5450}{12000},
        xmin=0, xmax=6.28,
        xtick={0, 1.57, 3.14, 4.71, 6.28}, 
        xticklabels={$0$, $\frac{\pi}{2}$, $\pi$, $\frac{3\pi}{2}$, $2\pi$},
            height=4cm, width=\columnwidth]
        \addplot [no marks] table [x=pathparam, y=err, col sep=comma] {figs/non_quantized_NN-S.csv};
        \addlegendentry{DNN}
    \end{axis}
\end{tikzpicture}
\begin{tikzpicture}
    \begin{axis}
        [
        legend pos= south east,
            skip coords between index={5500}{12000},
        xmin=0, xmax=6.28,
        xtick={0, 1.57, 3.14, 4.71, 6.28}, 
        xticklabels={$0$, $\frac{\pi}{2}$, $\pi$, $\frac{3\pi}{2}$, $2\pi$},
            height=4cm, width=\columnwidth]
        \addplot [no marks] table [x=pathparam, y=err, col sep=comma] {figs/quantized_NN-S.csv};
        \addlegendentry{QDNN}
    \end{axis}
\end{tikzpicture}
\begin{tikzpicture}
    \begin{axis}
        [xlabel = $\theta$,
        legend pos= south east,
            skip coords between index={5030}{12000},
        xmin=0, xmax=6.28,
        xtick={0, 1.57, 3.14, 4.71, 6.28}, 
        xticklabels={$0$, $\frac{\pi}{2}$, $\pi$, $\frac{3\pi}{2}$, $2\pi$},
            height=4cm, width=\columnwidth]
        \addplot [no marks] table [x=pathparam, y=err, col sep=comma] {figs/quantized_NN-S_P.csv};
        \addlegendentry{QDNN+P}
    \end{axis}
\end{tikzpicture}
    \caption{Cartesian position error with respect to the reference path
    for different implementations. The MPFC solution as computed by OpEn is 
    the most accurate.  The next most accurate is the proposed approach 
    (QDNN+P).}
\label{fig:patherror}
\end{figure}
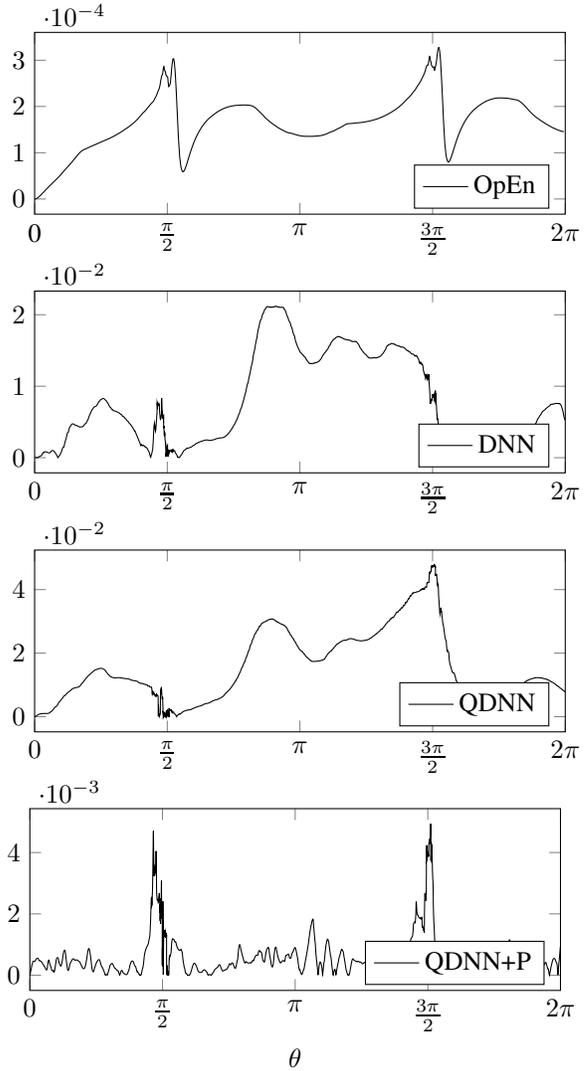

Fig. \ref{fig:path} shows a comparison of the path 
in the Cartesian $XY$ plane 
followed by the simulated robot 
using different implementations.
The absolute Cartesian position error is shown in Fig. \ref{fig:patherror},
with a summary presented in Table \ref{tbl:patherror}.
All implementations can follow the path, 
with OpEn being the most accurate.
When the worst-case error is considered, using a regular (non-quantized) 
DNN is two orders of magnitude worse than the OpEn implementation. 
The QDNN implementation has worse overall performance than the non-quantized network.
Finally, the proposed addition of two P controllers to the QDNN 
reduces its worst-case error by an order of magnitude and
outperforms the DNN.

\begin{table}
    \centering
    {\renewcommand{\arraystretch}{1.2}

\begin{tabular}[h]{|c|c|c|}
    \hline
    & Mean & Max.  \\
    \hline
    OpEn & 1.9E-4 & 3.3E-4 \\
    DNN & 7.5E-3 & 2.1E-2 \\
    QDNN & 1.6E-2 & 4.8E-2 \\
    QDNN+P & 6.1E-4 & 4.9E-3 \\
    \hline
\end{tabular}
    }
\caption{Mean and maximum Cartesian error in the path.}
\label{tbl:patherror}
\end{table}

\section{Conclusions}
\label{sec:concl}

The paper presented a model predictive path following implementation using
quantized deep neural networks augmented with a controller for quantization error compensation.
We showed a practical way to select the training set, and 
how to design the error compensation controller.
Compared to a traditional MPFC using online optimization,
our proposed approach requires only a fraction of the memory and runs several orders of 
magnitude faster on PC simulations.
Although the path-following accuracy is slightly degraded, we believe the performance may still be good for low-cost applications.
With a hardware-in-the-loop implementation using a microcontroller, we showed 
the effectiveness of this approach for low-cost embedded devices. 
Future work will discuss how to handle different path geometries with one trained QDNN and how to give performance guarantees.

\appendices 
\section*{Appendix}
The hyperparameters of the 
QDNN network are the learning rate $=4.5E-4$, 
the activation function (ReLU),
the number of hidden layers $H=9$, 
and the units on each layer:
input layer $4$ units, followed by the
hidden layers with $48$, $16$, $24$, $16$, $16$, $40$, $24$, $16$, and $24$ units, 
and output layer $3$ units.

The parameters of OCP \eqref{eqn:ocp} are 
the matrices
$Q=\textrm{diag}(2E5, 2E5, 1E5, 0)$, 
$R=\textrm{diag}(1E1, 5E3, 1E5)$, 
the box sets 
$\set{Z} = \{z \inRpow{4} \ | \  \underline{z} \leq z \leq \overline{z} \}$,
with 
$\underline{z} = [-5, -15, -\frac{\pi}{2}, -5]$,
$\overline{z} = [5, 15, \frac{\pi}{2}, -5]$, and 
$\set{W} = \{w \inRpow{3} \ | \ \underline{w} \leq w \leq \overline{w} \}$,
with 
$\underline{w} = [-0.26, -0.455, 0]$,
$\overline{w} = [0.26, 0.455, 0.15]$,
the discretization time $\delta=0.01$s,
and the horizon length steps $N=60$.

\bibliographystyle{IEEEtran}
\bibliography{IEEEabrv,literature}
\end{document}